\documentclass[aps,prb,twocolumn,superscriptaddress,showpacs,longbibliography,notitlepage]{revtex4-1}
\usepackage[colorlinks=true,citecolor=blue,linkcolor=blue,breaklinks=true]{hyperref}
\usepackage{amssymb}
\usepackage{graphicx}
\usepackage{amsmath}
\usepackage{mathrsfs}
\usepackage{mathtools}
\usepackage[export]{adjustbox}
\usepackage{epsfig}
\usepackage{times}
\usepackage{color}
\usepackage{setspace}
\usepackage{calc}
\usepackage{natbib}
\usepackage{bm}
\DeclareMathAlphabet{\mathpzc}{OT1}{pzc}{m}{it}
\begin{document}

\newcommand {\beq} {\begin{equation}}
\newcommand {\eeq} {\end{equation}}
\newcommand {\bqa} {\begin{eqnarray}}
\newcommand {\eqa} {\end{eqnarray}}
\newcommand {\ba} {\ensuremath{b^\dagger}}
\newcommand {\Ma} {\ensuremath{M^\dagger}}
\newcommand {\psia} {\ensuremath{\psi^\dagger}}
\newcommand {\psita} {\ensuremath{\tilde{\psi}^\dagger}}
\newcommand{\lp} {\ensuremath{{\lambda '}}}
\newcommand{\A} {\ensuremath{{\bf A}}}
\newcommand{\Q} {\ensuremath{{\bf Q}}}
\newcommand{\kk} {\ensuremath{{\bf k}}}
\newcommand{\qq} {\ensuremath{{\bf q}}}
\newcommand{\kp} {\ensuremath{{\bf k'}}}
\newcommand{\rr} {\ensuremath{{\bf r}}}
\newcommand{\rp} {\ensuremath{{\bf r'}}}
\newcommand {\ep} {\ensuremath{\epsilon}}
\newcommand{\nbr} {\ensuremath{\langle ij \rangle}}
\newcommand {\no} {\nonumber}
\newcommand{\up} {\ensuremath{\uparrow}}
\newcommand{\dn} {\ensuremath{\downarrow}}
\newcommand{\rcol} {\textcolor{red}}
\newcommand{\bcol} {\textcolor{black}}
\newcommand{\bu} {\bold{u}}
\newcommand{\tr} {\text{tr}}
\newcommand{\ac}[1]{\textcolor{magenta}{\textsf{ #1}}}

\begin{abstract}

 \end{abstract}
\title{Renyi Entropy of Interacting Thermal Bosons in Large $N$ Approximation}
\author{Ahana Chakraborty}\email{ahana@pks.mpg.de}
 \affiliation{Max Planck Institute for the Physics of Complex Systems, N\"othnitzer Str. 38, 01187, Dresden, Germany.}
\author{Rajdeep Sensarma}\email{sensarma@theory.tifr.res.in}
 \affiliation{Department of Theoretical Physics, Tata Institute of Fundamental
 Research, Mumbai 400005, India.}

\date{\today}

\begin{abstract}
  Using a Wigner function based approach, we study the Renyi entropy
  of a subsystem $A$ of a system of Bosons interacting with a local repulsive
  potential. The full system is assumed to be in thermal equilibrium
  at a temperature $T$ and density $\rho$. For a ${\cal U}(N)$ symmetric model, we show that the
  Renyi entropy of the system in the large $N$ limit can be understood
  in terms of an effective non-interacting system with a spatially
  varying mean field potential, which has to be determined self
  consistently. The Renyi entropy is the sum of two terms: (a) Renyi
  entropy of this effective system and (b) the difference in thermal
  free energy between the effective system and the original translation
  invariant system, scaled by $T$. We determine the self consistent equation for this
  effective potential within a saddle point approximation. We use this
  formalism to look at one and two dimensional Bose gases on a
  lattice. In both cases, the potential profile is that of a square
  well, taking one value in the subsystem $A$ and a different value outside
  it. The potential varies in space near the boundary of the subsystem $A$ on the scale of density-density correlation length. The effect of interaction on the
  entanglement entropy density is determined by the ratio of the
  potential barrier to the temperature and peaks at an intermediate
  temperature, while the high and low temperature regimes are
  dominated by the non-interacting answer.
  \end{abstract}

\maketitle

\section{Introduction}
Quantum entanglement and its various measures like entanglement
entropies have become an important theoretical tool in the study of
quantum many body systems. Starting from their domain of origin in the
area of quantum information~\cite{Entanglement_Review,Horodecki,NielsenChuang}, where
entanglement is considered as a resource for quantum computation~\cite{Ent_Resource},
these measures have played an important role in the study of systems
relevant to a wide variety of fields, from condensed matter theory to
quantum gravity~\cite{Ryu_2006}.

In condensed matter systems,
entanglement entropy has been used to detect phase
transitions~\cite{Vidal_Kitaev,Renyi_QPT2,Renyi_QPT3,melko1,melko2} and classify non-trivial
topology of ground state of quantum systems
~\cite{jiang2012topological,KitaevPreskillTopo,Tarun_Ashvin_topo,Levin_topo}. Entanglement entropy~\cite{EE_thermalization} and
entanglement spread~\cite{OTOC, Ent_Spread} has been widely used to understand thermalization
of complex many body systems. It has also been used to study many body
localization~\cite{MBLReview,MBLReview2,samanta2020extremal}, where strongly disordered interacting quantum systems do
not thermalize at long times. Entanglement entropy of many
body system has recently been measured in experiments with ultracold atomic systems~\cite{Islam_expt}.

While there is a large body of literature about entanglement entropy
of Fermionic~\cite{Fermion_EE1,Fermion_EE2} and spin systems~\cite{Calabrese_2005}, relatively less attention has been paid
to microscopic calculations of entanglement entropy of interacting Bosons. This is mainly due to the fact that
the larger Hilbert space for Bosons do not allow numerical evaluation
of entanglement entropy, unless the system is in the Tonk's gas limit~\cite{Tonksgas,Tonksgas1}
in one dimension, or deep in the Mott insulator regime~\cite{MI_Ent,Krishnendu_EE}, where the
Hilbert space dimension can be manageable. The analytical approach
comes from field theoretic considerations, where one either uses a
replica based construction with a complicated manifold~\cite{Casini_Ent_QFT}, or takes
advantage of simplifications from conformal field theories (CFT) if one is
interested in a critical point~\cite{Cardy_Ent_CFT,SubirON}. While the CFT based
approaches are most effective in one spatial dimensions, recent work
has extended these considerations to more than one
dimensions~\cite{SubirWitczakKrempa} for critical Bosons.

In a recent work \cite{Chakraborty_WF} we have developed a new field
theoretic method, to calculate the Renyi entanglement entropy for a
generic (interacting) open or closed many body Bosonic system
undergoing arbitrary non-equilibrium dynamics \cite{AhanaRajdeepArbitInit}. This method calculates
the Wigner characteristic function~\cite{GlauberCahill} of a density matrix as the
Schwinger-Keldysh (SK) partition function~\cite{Kamenev} of the system in the
presence of sources turned on only at the time of measurement. For
reduced density matrix of a subsystem, the sources need to be turned
on only for degrees of freedom residing in the subsystem. The key new
development is that unlike the well known replica based field theoretic approach to
calculating entangelement entropy~\cite{Casini_Ent_QFT}, here one does not need to use
complicated manifolds and can
work with the standard field theory and its correlation functions. As
a result, one is not restricted to particular geometries for the subsystem.

In this paper, we use this technique to
calculate Renyi entanglement entropy of a thermal system of lattice
Bosons at a fixed density, interacting with a local Hubbard repulsion. We consider $N$
species of Bosons and a ${\cal U}(N)$ symmetric interaction between
the species~\cite{LargeNReview}. We show that in the large $N$ limit, the Renyi entropy of
a subsystem is composed of two terms: (a) The Renyi entropy of an effective
non-interacting system which has an additional spatially varying potential and (b) The difference in re-scaled thermal free energy
between this effective system and the original translation invariant
system within mean field approximation. This spatially dependent effective potential occurs due to the
presence of the interactions between the Bosons and has no analogy in
the case of non-interacting Bosons. Developing a systematic large $N$ expansion
for the Renyi entropy of the interacting theory and interpreting it in terms of an
effective non-interacting system with a self consistent spatially
dependent potential is the key new result of this paper.

We divide a system of Bosons into two mutually exclusive sub-systems $A$ and $B$.
We use field theoretic methods for calculating Wigner
functions to construct a functional integral for the 2nd Renyi
entropy of the subsystem $A$. We use saddle point approximation in the
large $N$ limit to derive a self-consistent equation for the effective
potential imposed by inter-particle interactions between the Bosons. We solve this equations numerically for a one dimensional
chain and a two dimensional square lattice with nearest neighbour
hopping and hence calculate the Renyi entropy of these systems. For
the effective potential profile, we
find (i) the potential profile takes the form of a step potential,
which is flat in the bulk of the subsystem
$A$ and the subsystem $B$ with different values and varies sharply
across the entanglement cut (ii) The potential in the subsystem $B$
is just the thermal Hartree shift of the original system. (iii) The
potential barrier $\Delta V$ increases with temperature and saturates at large
temperatures. The effect of interactions on the entanglement entropy
is controlled by $\Delta V/T$, which shows a peak at a characteristic
temperature. (iv) The
potential profile varies near the boundary of the subsystem $A$ on a
lengthscale given by the density-density correlation length of the
original thermal system. This variation would become important near a
phase transition where the correlation length diverges.

Focussing on the Renyi entropy, we find that the entropy scales
linearly with the size of the subsystem (as is expected for a thermal
system), which allows us to define an entanglement entropy density of
the subsystem. The entropy density increases monotonically with
temperature and saturates to its non interacting value at high
temperatures. The high and low temperature values are governed by the
density of the system. The effects of the interaction parameter is
largest at an intermediate temperature scale where the ratio of the
potential barrier to the temperature is largest. We also define an
excess entropy density due to tracing, which measures the additional
entropy per site of the subsystem relative to the entropy per
site of the thermal density matrix of the whole system. This excess
entropy, which indicates the additional randomness generated by tracing
over the subsystem $B$,
monotonically decreases with temperature and goes to zero at large
temperatures. We find these trends in both one dimensional and two
dimensional systems with quantitative differences. We would like to note that a microscopic calculation of Renyi entropy in more than one dimension is less explored in the literature and our field theory formalism, which is agnostic about the dimensionality, is a new step in this direction. 

We will now present a roadmap for navigating the paper. In Section
~\ref{WCFS2}, we review the well known relations between the Wigner
characteristic function (WCF) of a density matrix and the second Renyi
entropy of a subsystem. In Section~\ref{S2thermal}, we show that the WCF of
a reduced density matrix of a thermal system can be written as a
partition function with a specific set of sources within an imaginary time
field theory. We work out the case of a system of non-interacting
Bosons as an example in this section. In Section~\ref{UNthermo}, we define
the ${\cal U}(N)$ symmetric interacting model of $N$ species of Bosons
and review its equilibrium field theory in the large $N$ limit. In
Section~\ref{S2largeN}, we focus on calculating the Renyi entropy of 
the ${\cal U}(N)$ symmetric model in the large $N$ limit. The key
results and the interpretation in terms of an effective potential is
developed in this section. In Section~\ref{Renyi1d}, we apply this formalism
to study  a one dimensional Bose gas, while in
Section~\ref{Renyi2d}, we look at the case of a two dimensional Bose
gas. We conclude in Section~\ref{Summary} with a summary and possible
future directions to extend this formalism.

\section{Wigner Functions and Renyi entropies \label{WCFS2}}

The Wigner function of a density matrix~\cite{Wigner1932} is the closest equivalent to a
phase space distribution function for a quantum system. For a many
body bosonic system, it is useful to define the Wigner Characteristics function (WCF), $\chi_W$ of the density matrix $\hat{\rho}$~\cite{GlauberCahill},
\begin{eqnarray}
\label{chiW_def}
\chi_W(\{\gamma_r\})&=& \textrm{Tr} \left[\hat{\rho} \hat{D}(\{\gamma_r\} )\right]=Tr \left[\hat{\rho}~ e^{\sum \limits_{r} \gamma_r c^\dagger_r-\gamma^\ast_r c_r}\right],
\end{eqnarray}
where $c^\dagger_r$ is the Bosonic creation operator for the spatial
coordinate $r$, and $\hat{D}$ is the displacement operator which
creates coherent states by acting on the ground state. Note that one
can use any complete single particle basis to define the Wigner
characteristics; we choose the position basis since it is the natural
choice for calculation of entanglement entropies. The Wigner
function is then defined by
\begin{eqnarray}
\label{W_def}
W(\{\alpha_r\})&=& \left[\int\prod_r \frac{d^2\gamma_r}{\pi^\Omega} e^{\sum
                   \limits_{r} \gamma^\ast_r \alpha_r-\gamma_r
                   \alpha^\ast_r}\chi_W(\{\gamma_r\}) \right],
\end{eqnarray}
where we consider a system with $\Omega$ lattice sites. One can follow the
same procedure, with the density matrix replaced by the operator
$\hat{O}$, to obtain the Weyl symbol of the operator
\begin{eqnarray}
\label{WeylA_def}
\chi_O(\{\gamma_r\})&=& \textrm{Tr} \left[\hat{O} \hat{D}(\{\gamma_r\} )\right]
  \\
 \nonumber W_O(\{\alpha_r\})&=& \left[\int\prod_r \frac{d^2\gamma_r}{\pi^\Omega} e^{\sum
                   \limits_{r} \gamma^\ast_r \alpha_r-\gamma_r
                   \alpha^\ast_r}\chi_O(\{\gamma_r\}) \right].
\end{eqnarray}
The expectation of the operator $\hat{O}$ is then given by the phase
space integral
\begin{eqnarray}
\label{expectation}
\nonumber \langle \hat{O}\rangle &=& \textrm{Tr} ~\hat{\rho}\hat{O}=\left[\int\prod_r \frac{d^2\alpha_r}{\pi^\Omega}
                           W(\{\alpha_r\}) W_O(\{\alpha_r\})\right]\\
  &=&\left[\int\prod_r \frac{d^2\gamma_r}{\pi^\Omega}\chi_W(\{\gamma_r\}) \chi_O(\{-\gamma_r\})\right]
\end{eqnarray}
To calculate entanglement measures in a system, one considers a
subsystem $A$ consisting of $\Omega_A$ sites. The reduced density matrix of
the sub-system $A$, i.e. $\hat{\rho}_{A}(t)=Tr_B \hat{\rho}(t)$ is
obtained by integrating out the degrees of freedom residing outside
$A$  (i.e. in  $B=A^c$ ). The second Renyi entropy of the
subsystem $A$ is then given by 
\beq
S^{(2)}= -\ln \textrm{Tr} ~\hat{\rho}_A^2.
\label{S2def}
\eeq
The Wigner characteristic function of the reduced density matrix can
be obtained from the full density matrix in the following way,
\begin{eqnarray}
\label{chiWA_def}
\chi^A_W(\{\gamma_j\})&=& \textrm{Tr} \left[\hat{\rho} \hat{D}(\{\gamma_j\} )\right]=\textrm{Tr} \left[\hat{\rho}~ e^{\sum \limits_{j} \gamma_j c^\dagger_j-\gamma^\ast_j c_j}\right],
\end{eqnarray}
where the lattice index $j$ runs only over the $\Omega_A$ sites in the
subsystem $A$, whereas the trace is over all degrees of freedom. The
trace over the degrees of freedom in $B$ creates the reduced density
matrix, while the displacement operator with support only in $A$ leads
to the Wigner characteristic function of the reduced density
matrix. Thus the WCF of the reduced density matrix is obtained from
Eq.~\ref{chiW_def} by restricting the support of the displacement
operator to the Hilbert space of the subsystem $A$. For the sake of simplicity and
clarity of the notation, throughout this paper, we denote the lattice
sites belonging to the sub-system, $A$ by the indices $j,j'$ etc, i.e. $j,j' \in A$ can run
only over the $\Omega_A$ lattice sites, whereas we will use $r,r'$ (or $x,y$) to
indicate co-ordinates that run over the full system with $\Omega$ sites.

Finally, using Eq.~\ref{S2def} together with Eq.~\ref{expectation}, we
can write the second Renyi entropy of the subsystem in terms of the
WCF of the reduced density matrix as 
\begin{equation}
\label{S2_def_chiA}
S^{(2)} =- \ln \left[\int\prod_{j\in A} \frac{d^2\gamma_j}{\pi^{\Omega_A}}|\chi^{A}_W( \{ \gamma_j\},t)|^2\right] .
\end{equation}
We note that these results are quite general and are applicable to
both non-interacting and interacting systems in and out of thermal
equilibrium. In Ref.~\onlinecite{Chakraborty_WF}, a general
Keldysh theory based formalism for calculating Renyi entropies of
systems out of equilibrium has been formulated. However, in this
paper, we focus on
Renyi entropy of systems in thermal equilibrium and provide a detailed
blueprint for calculating it within a field theory formalism. In the
next section, we relate the Wigner characteristic of the reduced
density matrix of a thermal system to a partition function with
appropriate source configurations and apply it to a system of
non-interacting Bosons.

\section{Renyi Entropy of Thermal Systems\label{S2thermal}}

For a system in thermal equilibrium, the density matrix in the grand
canonical ensemble is given by
$\hat{\rho} = \frac{1}{Z}e^{-\beta(\hat{H}-\mu\hat{N})}$, where $\hat{H}$ is the
Hamiltonian of the system, $\beta=1/T$  is the inverse temperature, $\hat{N}$ is the
total number operator and $\mu$ is the chemical potential. The
partition function is given by $Z=
\textrm{Tr}~e^{-\beta(\hat{H}-\mu\hat{N})}$, which ensures
$\textrm{Tr}~\hat{\rho}=1$. The partition function can be expanded in
a standard functional integral
\bqa
Z&=& \int D[\phi] e^{-{\cal A}(\phi^\ast_r,\phi_r)},\\
\nonumber {\cal A}(\phi^\ast_r,\phi_r)&=&\! \! \! \int _0^\beta \! \! d\tau \! \! ~ \sum_r \!
\phi^\ast_r(\tau)(\partial_\tau -\mu)\phi_r(\tau) \! +\! H [\phi^\ast_r(\tau),\phi_r(\tau)]
\eqa
where ${\cal A}$ is imaginary time action and
$\hat{H}[c^\dagger_r,c_r]$ is the normal ordered Hamiltonian of the
system, and $\phi_r(\tau)$ are the Bosonic fields.

The WCF of the reduced density matrix of the subsystem $A$ is given by
\bqa
&& \chi_W^A(\{\gamma_j\})= \frac{1}{Z} \textrm{Tr}~\left[ e^{-\beta(\hat{H}-\mu\hat{N})} \times e^{\sum_{j } \gamma_j c^\dagger_j 
  -\gamma^\ast_j c_j} \right] \\
\nonumber &=&\frac{1}{Z} \textrm{Tr}~\left[e^{-(\beta-\beta_0)(\hat{H}-\mu\hat{N})} \times \hat{D}(\{\gamma_j\})e^{-\beta_0(\hat{H}-\mu\hat{N})} \right]
 \label{chi_therm}
\eqa
where $\beta_0$ is an arbitrary number between $0$ and $\beta$, and we
have used the cyclic properties of the trace to write the last
line. The numerator can be expanded in a functional integral
\bqa
\chi_W^A(\{\gamma_j\})&=&\frac{1}{ Z } \int D[\phi]
~e^{-\mathcal{A}[\phi,\phi^*] - \left( \sum \limits_{j }\gamma^*_j \phi_j(\beta_0)-h.c.\right)}. ~~
\label{chiW_pathint}
\eqa
The
insertion of the displacement operator at the imaginary time
$\tau_0=\beta_0$ is equivalent to turning on a source~\footnote{Note that we have chosen here a slightly non-standard definition of partition function with linear sources compared to that in the standard thermal field theory, $Z[J_j(\tau)]=\int D[\phi] exp[-\mathcal{A}(\phi,\phi^*)+\mathbf{i} \{ \int_0^{\beta}J^*_j(\tau)\phi_j(\tau)+J_j(\tau)\phi^*_j(\tau)\} ]$. This definition together with the choice $J_j(\tau)=-\mathbf{i} \gamma_j \delta(\tau-\beta_0)$ takes care of
the relative negative sign in $\hat{D}=exp[\sum_{j } \gamma_j
 a^\dagger_j  -\gamma^\ast_j a_j]$.} 
$-\mathbf{i} \gamma_j$ coupled to the
field $\phi^\ast_j(\beta_0)$ and $\mathbf{i} \gamma^\ast_j$ coupled to
$\phi_j(\beta_0)$. Note that the sources are turned on only for the
fields in the subsystem $A$. They are turned on at the time $\beta_0$
and turned off immediately thereafter, creating a delta function
profile in imaginary time. 
This suggests that the WCF $\chi_W^A(\{\gamma_j\})$ can be related to
the equilibrium partition function  \cite{matsubara} $Z$ in presence
of linear delta function sources, $J_j(\tau)$, turned on only at
$\tau=\beta_0$ for the modes $j \in A$, i.e.
\beq
\chi_W^A(\{\gamma_j\}) = \frac{1}{Z}Z[J_j(\tau)=-\mathbf{i} \gamma_j \delta(\tau-\beta_0)], ~\forall j \in A.
\eeq
where the denominator is calculated without any additional
sources.
This identification matches earlier results presented in
Ref.\onlinecite{Chakraborty_WF}, and can be considered as a
special case of that more general result defined on real time Keldysh contours. Finally we note that the
choice of $\beta_0$ is completely arbitrary and hence
$\chi_W^A(\{\gamma_j\})$ is independent of the value of the imaginary
time when the source is turned on. This will play a very important
role when we consider an interacting model of Bosons in the next section.

We first apply this technique of calculating WCF and
Renyi entropy of Bosonic systems to the simplest case of a generic
non-interacting system of Bosons, with Hamiltonian, $H_0=\sum_{p}
\epsilon_{p} c^{\dagger}_{p} c_{p} $ (See Ref.~\onlinecite{Drut_2017}
for an alternate approach). We assume that the
single-particle eigenstate $p$ has a wavefunction $\psi_p(r)$. We
consider Bosons with  density $\rho=(1/\Omega)\sum_{p} n_B(\epsilon_{p}-\mu,T)$,
where $n_B(e,T)=1/[exp(e/T)-1]$, is the Bose distribution function and
the chemical potential $\mu(T)$ is tuned to obtain the correct density
at all temperatures. In this
case, the action ${\cal A}$ is quadratic and we get 
%
%
%
\bqa
\nonumber
\mathcal{A}[\phi,\phi^*] &=& \int \limits_{0}^{\beta}\int \limits_{0}^{\beta} d\tau d\tau' \sum \limits_{r,r'}\phi^\ast_r(\tau) G_0^{-1}(r,\tau;r',\tau') \phi_r(\tau').
\eqa
Here, $G_0^{-1}(r,\tau;r',\tau')$ is the imaginary time inverse Green's function of the non-interacting system. Performing the Gaussian integration over the fields $\phi_r(\tau)$ in Eq.\ref{chiW_pathint}, we obtain
\beq
\chi_W^A(\{\gamma_j\}) = e^{  -\sum \limits_{j,j' } \gamma^*_j M^A(j,j')  \gamma_{j'} },
\eeq
where, $M^A(j,j')= G_0(j,\beta_0;j',\beta_0)$ is a $\Omega_A \times
\Omega_A$ matrix. We note that one should be careful in taking the equal
time limit $\tau=\tau'=\beta_0$ in $\hat{M}^A$ 
from the time ordered Green's function, $G_0(j,j';\tau-\tau')$,
symmetrically along the imaginary time axis; i.e. 
\bqa
\label{M_eqtime}
M^{ A}(j,j')&=& \frac{1}{2} \lim_{\eta \rightarrow 0^+} \Big[  G_0(j,\beta_0;j',\beta_0+\eta) \nonumber \\&&~~~~~~~~~~~~  + G_0(j,\beta_0;j',\beta_0-\eta) \Big].
\eqa
The symmetric limit is due to the fact that the Wigner function is
related to a symmetric rather than a (anti) normal ordering of $c$ and
$c^\dagger$ operators. This limit also matches with the more general
answer derived in Ref.~\onlinecite{Chakraborty_WF} using Keldysh field
theory. Note that
in this case the answer is independent of the choice of $\beta_0$, as
we had predicted before on very general grounds. The system is still time-translation invariant along the imaginary time axis, $\tau$ and we can obtain the imaginary time Green's functions by the standard Matsubara summation technique\cite{matsubara}. This yields,
\beq
M^A(j,j') = \frac{1}{2}\sum \limits_{p} \psi_{p}(j)\psi^{*}_{p}(j') \left[ 1+2n_B(\epsilon_{p}-\mu,T) \right],
\eeq
We now obtain $S^{(2)}$ by performing the Gaussian integration over the $\gamma_j$ variables in Eq. \ref{S2_def_chiA},
\beq
S^{(2)} =Tr_A \ln \left( 2 \hat{M}^A \right).
\label{S2_NI}
\eeq
The entanglement entropy scales linearly with the subsystem size
$\Omega_A$, as expected for a thermal density matrix. 
The above expression is further simplified if we calculate the Renyi
entropy of the full system i.e. $j$ runs over all the
lattice sites. In this case, $\hat{M}^A$ is a $\Omega \times \Omega$ matrix with eigenvalues, $m_{p}=[ 1+2n_B(\epsilon_{p}-\mu,T)]/2$. Substituting this in Eq. \ref{S2_NI}, we get
\beq
S^{(2)} = \sum \limits_{p} \log \left[ \coth \left(
    \frac{\epsilon_{p}-\mu}{2 T} \right) \right].
\label{S2_ni}
\eeq
%
The above example of a simple case of non-interacting Bosonic systems
gives a consistency check of the new field theoretic method proposed
to calculate $S^{(2)}$~\cite{Drut_2017}. In the subsequent sections,
we will use this technique to calculate Renyi entropy of a reduced
density matrix in presence of inter-particle interactions in the
system.

\section{$\mathcal{U}(N)$ model and large-N field theory\label{UNthermo}}

In the previous section, we have obtained exact expressions  for
entanglement entropy of a subsystem of non-interacting Bosons which
are in thermal equilibrium. Our aim is to extend this formalism to the
case of interacting Bosons using a large $N$ formulation.  To this
end, we consider a Bose Hubbard model with $N$ species of Bosons. The
Bosons hop between nearest-neighbour sites on a lattice and interact with local repulsive inter-species interaction given by,
\beq
\label{ham_interacting}
H=-t\sum \limits_{\langle r r' \rangle} \sum \limits_{a=1}^{N}c^{\dagger(a)}_r c^{(a)}_{r'} + \frac{U}{2N} \sum \limits_{r} \sum_{a,b=1}^{N}c^{\dagger(a)}_rc^{\dagger(b)}_r c^{(a)}_rc^{(b)}_r
\eeq
where $\langle r,r' \rangle$ denote the nearest-neighbour lattice sites, $a,b$ are species indices which run
from $1$ to $N$, $t$ is the
hopping matrix element and $U$ is the scale of the local repulsion. We
note that the scaling of the interaction, $U/N$, ensures that the energy
of the system is extensive in the number of species $N$. It can be
easily seen that this
Hamiltonian is invariant under a global unitary rotation between the
species, $c^{(a)}_r \rightarrow \tilde{c}^{(a)}_r=\mathcal{U}_{ab}
c^{(b)}_r$, where $\mathcal{U}$ is an arbitrary $N \times N$ unitary
matrix. We would study this $\mathcal{U}(N)$ symmetric model in the
limit of large $N$.

Although our final aim is to calculate entanglement entropy of
subsystems in this model, in this section, we will focus on the
thermodynamics of the model. We will now calculate the partition function and
hence free energy of this interacting Bosonic system in thermal
equilibrium, using a large $N$ approximation~\cite{LargeNReview}. As we will find in the
next section, this will form a part of the calculation of Renyi
entropy of the subsystems. This will also allow us to discuss the
large $N$ approximation and resulting saddle points in a known and
simple example~\cite{LargeNReview}, while setting up notation for the large $N$
calculation of entanglement entropy, which we will take up in the next
section. 

For the standard thermal field theory, the
  action $\mathcal{A}$ of this $\mathcal{U}(N)$ symmetric model  is
  given by
\begin{widetext}
\bqa
\label{Action_LargeN}
\mathcal{A}[\phi,\phi^*] \! &=& \!  \sum_{a=1}^{N}\! \int_0^\beta \! \! \! d\tau \! \sum
\limits_{r} \phi^{*(a)}_r(\tau) \! [\partial_\tau
-\mu]\phi^{(a)}_r(\tau)\! -\! t \! \sum\limits_{\langle rr'\rangle} \phi^{*(a)}_r(\tau)\phi^{(a)}_{r'}(\tau) 
\!+\! \frac{U}{2N}\! \sum_{a,b=1}^{N} \! \int_0^\beta \! \! \! d\tau \! \sum \limits_{r} \! \phi^{*(a)}_r(\tau)\phi^{*(b)}_r(\tau)\phi^{(a)}_r(\tau)\phi^{(b)}_r(\tau).~~~~~
\eqa
%
Here, $\mu=\mu(T)$ is a
chemical potential which ensures the correct number density per
species $\rho$ in the
system. We will vary $\mu$ so that $\rho$ is kept fixed at all temperatures.
To calculate the partition function of this system in the large $N$
limit,  we perform the standard Hubbard Stratonovich (HS) transformation on the quartic interaction term by introducing the auxiliary fields, $\lambda_r(\tau)$, which couples to the bi-linears of the fields, $\sum_{a=1}^{N}\phi^{*(a)}_r(\tau)\phi^{(a)}_r(\tau)$. 
Then the partition function reduces to
%
\bqa
Z &=& \int D[\phi] D[\lambda] ~exp\left[-\frac{N}{2U}\int d\tau \sum \limits_{r} \lambda_r^2(\tau)-\int \limits_{0}^{\beta} \int \limits_{0}^{\beta} d\tau d\tau'  \sum \limits_{r,r'} \sum \limits_{a=1}^{N} \phi^{*(a)}_r(\tau) \left[ G^{-1}_{\lambda}(r,\tau;r'\tau')\right] \phi^{*(a)}_{r'}(\tau') \right],
\label{largeNaction}
\eqa
\end{widetext}
where the inverse propagator
\beq
\label{Int_Prop}
G^{-1}_{\lambda}(r,\tau;r'\tau')=G_0^{-1}(r,\tau;r',\tau') -\mathbf{i}\lambda_r(\tau)
\delta(\tau-\tau')\delta_{rr'}.
\eeq
Here the non-interacting inverse
propagator $G_0^{-1}(r,\tau;r',\tau')=\delta(\tau-\tau')\left[
  (\partial_\tau -\mu )\delta_{rr'}-t \delta_{\langle
    rr'\rangle}\right]$. The $\phi$ integrations
in Eq.~\ref{largeNaction} can be performed exactly to obtain,
\bqa
Z &=&\int D[\lambda] ~e^{-N~Tr \ln \left[ \hat{G}^{-1}_{\lambda}\right]  
-\frac{N}{2U}\int_0^\beta d\tau \sum \limits_{r} \lambda_r^2(\tau)  },
\label{norm_gauss}
\eqa
Note that each term in the exponent of
Eq.~\ref{norm_gauss} is multiplied by $N$. Now if we take the limit,
$N \rightarrow \infty$, we can evaluate the integral over $\lambda$ in
Eq. \ref{norm_gauss} in the saddle point approximation. Choosing a
static uniform saddle point solution $-\mathbf{i}\lambda_r(\tau)=\lambda_{th}$ , the
saddle point equation can be written as
\beq
\lambda_{th}=\frac{U}{\Omega}\sum \limits_{k} n_B(\epsilon_k+\lambda_{th}-\mu,T),
\label{saddlelambda}
\eeq
where the dispersion for a hypercubic lattice in $d$ dimension is
$\epsilon_k = -2t \sum_{i=1}^d \cos k_i$. In
addition, we have to solve the number equation
\beq
\rho=\frac{1}{\Omega}\sum \limits_{k} n_B(\epsilon_k+\lambda_{th}-\mu,T).
\label{numeqn}
\eeq
It is then easy to show
that the saddle point solution is $\lambda_{th}=U\rho$, which is the
standard Hartree energy shift in the system.

Substituting this saddle point value of the auxiliary variable
$\lambda_{th}$ in Eq. \ref{norm_gauss} we obtain the free energy within
the large $N$ approximation,
\beq
 F= \frac{N\Omega}{2 } U\rho^2 - NT\sum \limits_{k} \ln \left[ 1-e^{-\frac{\epsilon_k+\lambda_{th}-\mu}{T}} \right]
\eeq
We note that for stability of the system, the
effective chemical potential $\mu-\lambda_{th}$ must be lower than the
bottom of the band dispersion, $-2td$. Bose Einstein condensation in
this system occurs when $\mu-\lambda_{th}$ reaches the bottom of the
band. In one dimension, this happens only at $T=0$. In two dimensions
the large $N$ mean field theory for Schrodinger Bosons precludes the
possibility of Bose Einstein condensation due to a weak logarithmic
infrared divergence in the number equation. The physics of vortex
binding and unbinding, which is missing from this description, leads to
a BKT transition in this case. In three dimension, the large $N$
theory predicts a finite temperature phase transition. In this paper
we will confine ourselves to the ``high temperature'' or non-condensed
phases of the Bosons. The case of the ordered phase will be taken up
in a future work.

In the next section, we will modify the large $N$ procedure described
above for calculating the Wigner characteristic and hence Renyi
entropies of a subsystem of interacting thermal Bosons.


\section{Renyi entropy for $\mathcal{U}(N)$ model\label{S2largeN}}

In this section, we will devise a large $N$ approximation to calculate
the Renyi entropy of a subsystem of interacting Bosons. The standard large $N$ approximation for the thermal field theory of
$\mathcal{U}(N)$ model has two parts: a Hubbard Stratanovich
transformation with an auxiliary field coupled to a $\mathcal{U}(N)$
invariant bilinear of the fields, and a saddle point evaluation of the
integrals over the auxiliary fields, justified by the $N \rightarrow
\infty$ limit. Since the large $N$ approximation for the partition
function has already been shown in the last section, here we focus on
the partition function in presence of the sources, i.e. the WCF $\chi_W^A(\{\gamma_j\}) $.

We will use the  path integral representation of
$\chi_W^A(\{\gamma_j\}) $ shown in Eq. \ref{chiW_pathint}, with the
action given by Eq. \ref{Action_LargeN}. Note that in this case each
species of the Boson fields are coupled to sources. Following the steps in the
previous section, we will first decouple the interaction terms in
$\mathcal{A}$ by a HS transformation to get
\begin{widetext}
\bqa
\chi_W^{A}(\{\gamma_j\})&=&\frac{1}{Z } \int D[\lambda] D[\phi]  ~e^{\left[-\frac{N}{2U}\int d\tau \sum \limits_{r} \lambda_r^2(\tau)-\int \int d\tau d\tau' \sum \limits_{r,r'} \sum \limits_{a=1}^{N} \phi^{*(a)}_r(\tau) \left[ G^{-1}_{\lambda}(r,\tau;r'\tau')\right] \phi^{*(a)}_{r'}(\tau') \right]} \nonumber \\
&&~~~~~~~~~~~~~~~~~~~~~~~~~~~~~~~~~~~~~~~~~~~~\times e^{-\left( \sum \limits_{j } \sum \limits_{a=1}^{N} \gamma^{*(a)}_j \phi^{(a)}_j(\beta_0)-h.c.\right)}.~~~
\label{chi_W_UN_HS}
\eqa
where we would like to emphasize once again that $r,r'$ runs over all
lattice sites, while $j$ runs only over the subsystem $A$. The
partition function $Z$ in the denominator has already been calculated in the previous
section. The Gaussian functional integration over the fields $\phi_r(\tau)$ yields
\bqa
\chi_W^{A}(\{\gamma^{(a)}_j\})&=&\frac{1}{Z } \int D[\lambda] ~e^{-N~Tr \ln \left[ \hat{G}^{-1}_{\lambda}\right]  
-\frac{N}{2U}\int d\tau \sum \limits_{r} \lambda_r(\tau)^2 -  \sum \limits_{j,j'} \sum \limits_{a=1}^{N} \gamma^{*(a)}_j  M^A_{\lambda}(j,j') \gamma^{(a)}_{j'} },
\label{chi_D_UN}
\eqa
%
where, $\hat{G}^{-1}_\lambda$ is the interacting inverse propagator defined in
Eq.~\ref{Int_Prop}. Once again,
$M^A_{\lambda}(j,j')=G_{\lambda}(j,\beta_0;j',\beta_0)$, with the
equal time limit taken symmetrically as defined in Eq. \ref{M_eqtime}. In this case, it is useful to
substitute $\chi_W^A$ from Eq. \ref{chi_D_UN} in Eq. \ref{S2_def_chiA} and
perform the Gaussian integration over the source variables, $\gamma^{(a)}_j$.
%
%
We then obtain the Renyi entanglement entropy, $S^{(2)}$ from, 
%
\bqa
\nonumber
e^{-S^{(2)}}=\int\prod_{a, j
}\frac{d^2\gamma^{(a)}_j}{\pi^N}|\chi^{A}_W( \{ \gamma^{(a)}_j\})|^2
&=&\frac{1}{Z^2 } \int D[\lambda] \int D[\nu] e^{-N~\left\{Tr \ln \left[ \hat{G}^{-1}_{\lambda}\right]+Tr \ln \left[ \hat{G}^{-1}_{\nu}\right]  \right\} }\nonumber \\
&&~~~\times~
e^{-\frac{N}{2U}\int d\tau \sum \limits_{r}\left[ \{\lambda_r(\tau)\}^2+ \{\nu_r(\tau)\}^2\right] - N~ Tr_A \left[ \ln \{  \hat{M}^A_{\lambda}+ \hat{M}^A_{\nu} \} \right] }.
\label{S2LN}
\eqa
The exponent in the integral is multiplied by $N$, and in the limit of
large $N$, we can evaluate the above integral over the two auxiliary
fields using saddle point approximation. Note that the saddle point solutions will be different from that of the thermal field theory due to the extra term $\sim Tr_A  \ln \{  \hat{M}^A_{\lambda}+ \hat{M}^A_{\nu} \} $, which comes from integrating over the source variables $\gamma^{(a)}_j$ of the WCF. This term only depends on the Green's functions in the sub-system $A$.
Symmetry indicates that the
saddle point profiles $\lambda^{0}_r(\tau)=\nu^0_r(\tau)$. Considering
the saddle point condition, we get
\beq
\frac{\lambda^0_r(\tau)}{U} = \mathbf{i}G_{\lambda^0}(r,\tau ; r,\tau)-\frac{\mathbf{i}}{2}\sum \limits_{j,j'}~ \left[G_{\lambda^0}(r,\tau ; j,\beta_0) {M}_{\lambda^0}^{A(-1)} (j,j')  G_{\lambda^0}(j',\beta_0 ; r,\tau) \right].
\label{saddle}
\eeq
\end{widetext}
The first term in Eq.~\ref{saddle} is the standard answer one gets for
the thermodynamics of the system. This term is independent of $\tau$
as well as independent of $\beta_0$. The second term comes from
integrating out the sources $\gamma^{(a)}_j$. This is an additional
term that occurs when evaluating the Renyi entropy and depends explicitly on
$\tau$ as well as on $\beta_0$. However, as we had mentioned before,
the choice of $\beta_0$ is arbitrary and answers should not depend on
this variable. Hence, we integrate over the variable
$\beta_0$ from $0$ to $\beta$ and divide by $\beta$. This
approximation allows us to
consider a spatially varying, but (imaginary) time independent saddle point
profile $\lambda^0_r$, given by 
%
\begin{widetext}
\bqa
\frac{\lambda^0_r}{U}&=&  \mathbf{i}G_{\lambda^0}(r,\tau ; r,\tau) 
-\frac{\mathbf{i}}{2\beta}\int \limits_{0}^{\beta}  d\beta_0  \sum \limits_{j,j'}~ \left[G_{\lambda^0}(r,\tau ; j,\beta_0) {M}_{\lambda^0}^{-1} (j,j')  G_{\lambda^0}(j',\beta_0 ; r,\tau) \right] \nonumber \\
&=&  \mathbf{i}G_{\lambda^0}(r,r;0) 
-\frac{\mathbf{i}}{2\beta}\int \limits_{0}^{\beta}  d\tau  \sum \limits_{j,j'}~ \left[G_{\lambda^0}(r, j;-\tau) {M}_{\lambda^0}^{A(-1)} (j,j')  G_{\lambda^0} (j', r;\tau) \right].
\label{saddle_avg}
\eqa
\end{widetext}
Note that we have restored the (imaginary) time translation invariance
of the problem, and hence one can use Matsubara frequency sums to
evaluate the integrals above. The saddle point is then equivalent to
an effective non-interacting Hamiltonian with a spatially varying
external potential $V(r)=-\mathbf{i}\lambda^0_r$, which has to be
determined self consistently. If the eigenvalues and eigenfunctions of
this effective single particle Hamiltonian
\beq
H_{eff}=H_0 +V(r)-\mu,
\eeq
are $E_{n}$ and $\psi_{n}(r)$ ($H_0$ is the non-interacting part
of the original Hamiltonian given in Eq. \ref{ham_interacting}), the self-consistency equation is given by
\begin{widetext}
\bqa
\frac{V(r)}{U} &=& \sum \limits_n |\psi_n(r)|^2~n_B(E_n,T)+
T\sum \limits_{nn'} \sum \limits_{j,j' }  \psi_n(r)\psi^{*}_n(j)
\psi_{n'}(j')\psi^{*}_{n'}(r) \frac{ [ n_B(E_n,T) -n_B(E_{n'},T)
  ]}{E_{n}-E_{n'}}M_{\lambda^0}^{A(-1)}(j,j')
\label{saddleLambda}
\eqa
Substituting this saddle point solution, $V(r)$ in Eq. \ref{S2LN}, we
obtain the Renyi entanglement entropy in the large $N$ limit,
\bqa
\nonumber{\cal S}^{(2)}=\frac{S^{(2)}}{N} 
&=& -\frac{\beta}{U} \sum \limits_{r} V^2(r) +2 \sum \limits_{n} \ln
\left[ 1-e^{-\beta E_n} \right] +\frac{\Omega \beta}{U}  \lambda_{th}^2-2 \sum \limits_{p} \ln \left[
  1-e^{-\beta(\epsilon_p+\lambda_{th}-\mu)} \right].\\
&+&Tr_A \ln \left[ 2 \hat{M}^A_{V(r)} \right] 
\label{S2largeN}
\eqa
\end{widetext}
We note that the leading order approximation to the entanglement
entropy simply scales with the number of Boson components $N$, and
hence it is useful to define the entropy per species of the Bosons.
From the above formula we see that the entanglement entropy of the
system in large $N$ approximation has two contributions: (a) the first
line corresponds to the difference of the re-scaled thermodynamic free energy, $\beta F$ of two
systems: the original translation invariant system that we are
interested in, and the ``effective'' system with a spatially dependent
potential $V(r)$, stemming from the entanglement cut in the system,
which has to be self-consistently determined from
Eq.~\ref{saddleLambda} and (b) a second contribution (second line) related to the
one-particle correlation function $M$ within the subsystem $A$. This
correlation function is calculated for the system with the
``effective'' Hamiltonian. Note that in case of a non-interacting
system, there is no effective potential, and hence we simply get the
second line as the full answer. Our formalism thus generalizes the
non-interacting answers to the case of interacting system within the
large $N$ approximation.

In the next two sections, we will use this large $N$ solution to calculate the Renyi entropy of a $1$-d and $2$-d interacting Bose gas.

\section{Renyi entropy of 1D Bose gas\label{Renyi1d}}
\begin{figure*}
  \includegraphics[width=0.88\textwidth]{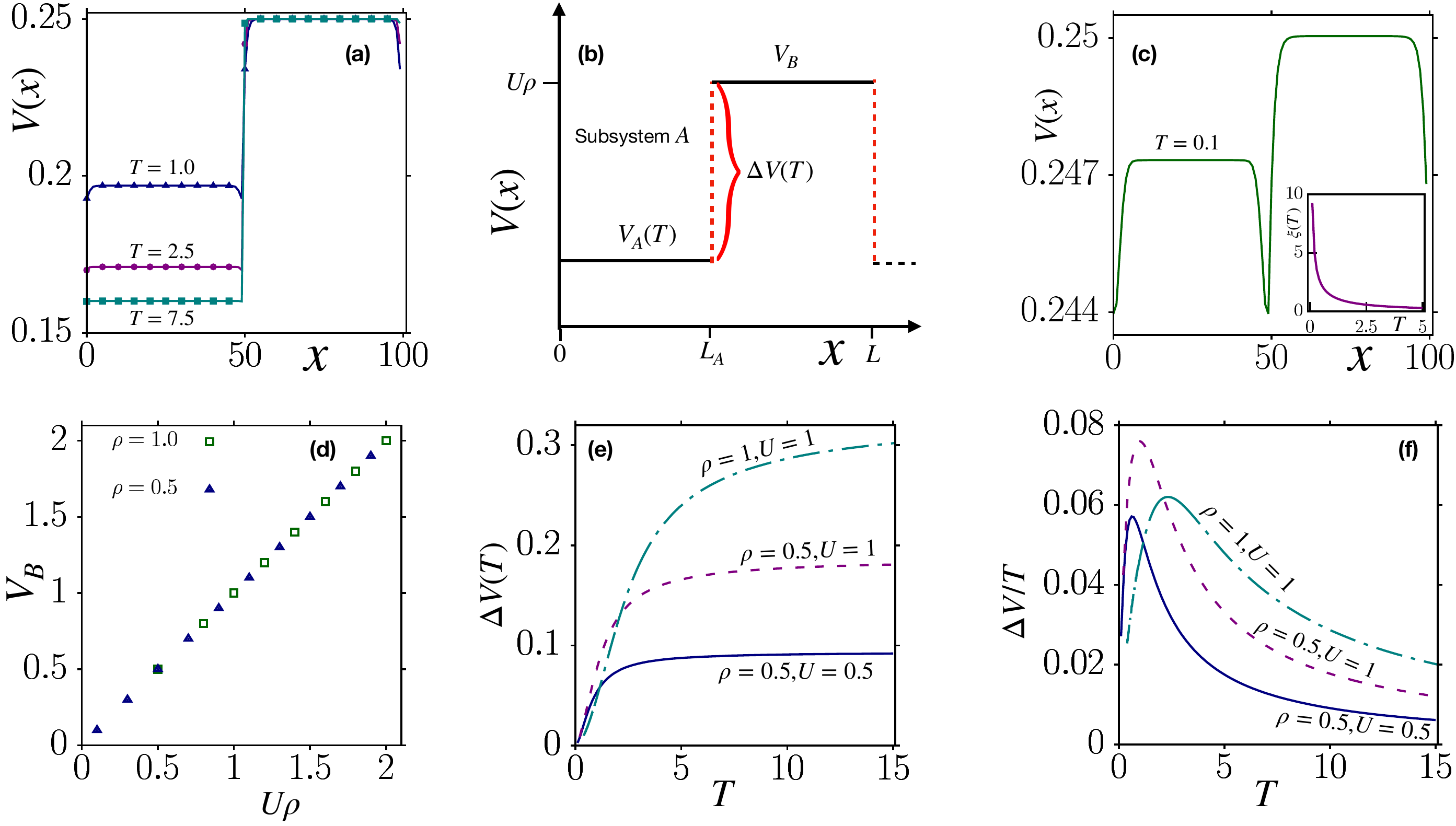}
  \caption{Profile of the effective potential $V(x)$ for 1D Bosons
    hopping on a $L=100$ site chain: The subsystem $A$ is the left
    half of the chain ($0 \leq x<50$), while the subsystem $B$,
    which is traced over, is on the right side ($50 \leq x<100$).
  (a) shows $V(x)$ vs $x$ for three different temperatures $T=7.5$
  (squares), $T=2.5$ (circles) and $T=1.0$ (triangles) at $U=0.5$ and
  $\rho=0.5$. At high temperatures, the profile closely resembles a step-potential with $V(x)=V_A(T)$ within the sub-system $A$ and $V(x)=V_B$ within the sub-system $B$. This profile is schematically shown in (b) where $\Delta V(T)=V_B(T)-V_A$ is height of the potential barrier at the entanglement cut ($x=L_A$). At low temperatures, this profile deviates from the step potential profile near the entanglement cut while in the bulk of the sub-systems $A$ and $B$, it has similar flat profile as shown in (c) for $T=0.1$. The lengthscale of this deviation across the boundary between $A$ and $B$ sub-systems closely matches with the connected density-density correlation length, $\xi(T)$ shown in the inset of the figure. At all temperatures $V_B$ is a temperature independent constant which is plotted as a function of $U\rho$ for two different densities, $\rho=0.5$ (triangles) and $\rho=1.0$ (squares) at $T=5.0$ in (d). The unit slope of the figure dictates that $V_B=U\rho$. (e) shows $\Delta V(T)$ as a function of $T$ for $\rho=0.5,U=0.5$ with solid line, $\rho=0.5,U=1$ with dashed line and $\rho=1,U=1$ with dashed-dotted line. The initial rise in $\Delta V(T)$ with increasing $T$ saturates at high $T$ where the saturation value increases with $U$ and $\rho$. (f) shows variation of  $\Delta V(T)/T$ with $T$ for the same parameters used in (e). $\Delta V(T)/T$ rises to a peak value at $T=T_{\mathrm{max}}$ before falling as $T^{-1}$ at high $T$. $T_{\mathrm{max}}$ increases with increasing $U$ and $\rho$. Since this dimensionless quantity $\Delta V(T)/T$ controls the effect of the inter-particle interaction in the thermal system, the effects of the latter in ${\cal S}^{(2)}$ will also be the largest at $T=T_{\mathrm{max}}$. In all the plots, we set $t=1$ and $a=1$ as units of energy and length respectively.
  }
  \label{1dBarrier}
  \end{figure*}
 We consider a one dimensional chain of size $L$, where Bosons hop
 with a nearest neighbour tunneling amplitude $t$ and interact locally
 with a repulsive interaction scale $U$. Considering $N$ species of Bosons, the ${\cal U}(N)$ symmetric Hamiltonian is given by
\bqa
H\! =\! \! -t \! \sum \limits_{ x } \! \sum \limits_{a=1}^{N} \! c^{\dagger(a)}_x c^{(a)}_{x+1}\! + \! h.c \!
+\!  \frac{U}{2N}\!  \sum \limits_{x} \! \sum_{a,b=1}^{N} \! c^{\dagger(a)}_xc^{\dagger(b)}_x c^{(a)}_xc^{(b)}_x,\nonumber
\eqa
where $a,b$ indicate the species index of the Bosons.
The Hamiltonian has periodic boundary conditions to ensure
translational invariance. This leads to a dispersion $\epsilon_p =-2t
\cos p$, where $p$ is the lattice momentum. We will set $t=1$ and the lattice-spacing $a=1$ to fix units of energy and length in all subsequent discussion in this paper. For 1D chain, the band energies vary between $-2$ and $+2$, giving a
bandwidth of $4t=4$. We consider the system at a temperature $T$
with a fixed density $\rho$, while we allow the chemical potential $\mu$ to
vary with temperature. This ensures the same number density at all temperatures.  We are
interested in the Renyi entropy of the reduced density matrix of a
contiguous subsystem of size $L_A$. Note that due to the translation invariance
of the original Hamiltonian, this subsystem can be placed anywhere in
the system; we choose to place it in the left half of the system. In
this section, we will work with $L= 100$, unless otherwise mentioned.
\begin{figure}
  \includegraphics[width=\columnwidth]{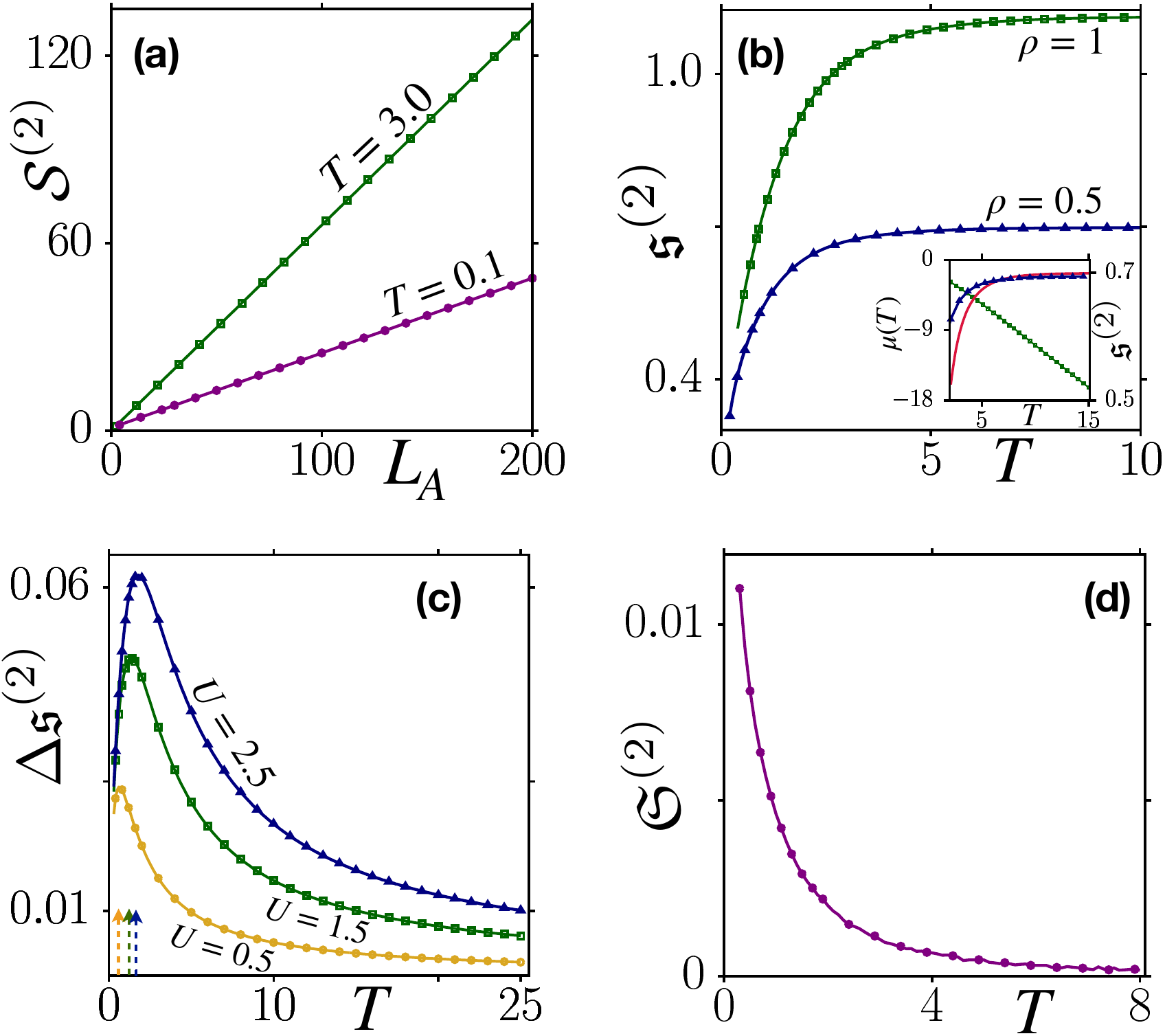}
  \caption{Renyi entropy of the reduced density matrix $\hat{\rho}^A$ consisting of the DOFs belonging to the sub-system $A$ ($0 \leq x<L_A$) for 1d Bosons: (a) shows ${\cal S}^{(2)}$ scales linearly with the sub-system length $L_A$ both at high temperature $T=3.0$ (squares) as well as in low temperature $T=0.1$ (circles).   We use $L=600$, $U=0.5$ and $\rho=0.5$ for this plot. The entanglement entropy density (entanglement entropy per site per
species of the 1D Bosons) $\mathfrak{s}^{(2)}$ is plotted as a
function of $T$ in (b) for two different value of densities,
$\rho=0.5$ (triangles) and $\rho=1.0$ (squares) for same value of
$U=1.0$. $\mathfrak{s}^{(2)}$ increases with $T$ and saturates at high
$T$, with the saturation value increasing with $\rho$. At large $T$, $\mathfrak{s}^{(2)}$ (triangles) closely agrees with the non-interacting answer plotted by red solid line in the inset for  $\rho=0.5$. This can be understood in the following way: the chemical potential $\mu(T)$ (solid line with squares) increases linearly with $T$ as shown in the inset of (b) and at large $T$, $\Delta V(T)$ imposed by the inter-particle interaction becomes negligible compared to $\mu(T)$. Hence, at large $T$ the effects of interaction becomes completely suppressed in $\mathfrak{s}^{(2)}$. To see this clearly, we plot $\Delta \mathfrak{s}^{(2)}=\mathfrak{s}^{(2)}(U)
-\mathfrak{s}^{(2)}(0)$ as a function of $T$ in (c) for three different strengths of interaction: $U=2.5$ (triangles) , $U=1.5$ (squares) and $U=0.5$(circles) for $\rho=0.5$. This shows that the effect of interaction in $\mathfrak{s}^{(2)}$ rises with $T$ and becomes maximum at an intermediate temperature scale which closely agrees with $T_{\mathrm{max}}$ (shown by three dashed vertical arrows) obtained from Fig. \ref{1dBarrier}(f). (d) shows monotonic decrease in additional entropy density $\mathfrak{S}^{(2)}=\mathfrak{s}^{(2)}(L_A)
-\mathfrak{s}^{(2)}(L)$ obtained from tracing out DOFs with increasing
$T$ for $U=1.0$ and $\rho=0.5$. We have used $L_A=50$ and $L=100$ for
(b) , (c) and (d). }
  \label{1dRenyi}
  \end{figure}

For a lattice
Boson system, the chemical potential lies below the bottom of the band
at high temperatures. If the chemical potential reaches the bottom of
the band, Bose Einstein condensation occurs in the system. It is
well known that in $1$ dimension, the low energy divergence of the single-particle
density of states $D(\epsilon)\sim \epsilon^{-1/2}$, together
with the $\sim \epsilon^{-1}$ divergence of the Bose function,
$n_B(\epsilon)=[e^{\beta\epsilon}-1]^{-1}$,  prevents the
formation of a BEC in the system. This remains true for the
interacting system within the large $N$ approximation. Thus our
formalism, which does not take into account possible Bose condensation, works at all
temperatures for the $1$ dimensional system.

The first task in calculating the Renyi entropy of a subsystem is to
obtain the self-consistent profile of the effective potential $V(x)$
from the saddle point equation Eq.~\ref{saddleLambda}. The
self-consistent profile obtained for three different temperatures,
$T=7.5$, $T=2.5$ and $T=1$ are plotted as a function of the lattice
sites in Fig.~\ref{1dBarrier}(a). In this case, $U=0.5$ and
$\rho=0.5$. From the figure, we see that the potential has a flat profile in
the bulk of subsystem $A$ with a value $V_A$ and a similar flat profile with a different
value $V_B$ in the bulk of
subsystem $B$. It changes near the entanglement cut at $x=50$ and
$x=0$ very rapidly over a scale of a few lattice spacings. This is
depicted schematically in Fig.~\ref{1dBarrier}(b) as a step potential barrier with a
height $\Delta V(T)=V_{B}(T)-V_A(T)$. Near the entanglement cut, we find that the potential in the subsystem $A$ drops near
the edge before rising sharply in the $B$ subsystem. The lengthscale
of variation on either side is almost equal to the correlation length
$\xi(T)$, obtained from the exponential decay of the connected
density-density correlation function in the thermal system. This is not surprising if we think of the entanglement cut
as a localized source of disturbance. One would then expect the local density, and hence the
effective potential, to adjust to this perturbation on a scale of $\xi(T)$. To see this clearly, we plot the potential profile at
$T=0.1 $ in Fig.~\ref{1dBarrier}(c), where the variation of the
potential occurs on a measurable length scale of $\sim 9$. In the inset of Fig.~\ref{1dBarrier}(c), we plot $\xi(T)$ with the temperature of the
system. We see that at $T=0.1$, $\xi(T)$ is also $ \sim 9 $. The key takeaway from this is that, while the
entanglement calculation inevitably involves solving a translation
non-invariant problem of the effective potential (even though the
original system is translation invariant), one can get away with
solving a much simpler problem of a step
potential well in the subsystem $A$, as long as the correlation
length $\xi(T)$ is much smaller than the subsystem size $L_A$. However, the value of the barrier needs to be solved self-consistently. This
can in principle be used to calculate entanglement entropies in large
systems in higher dimensions. We note that this potential barrier is imposed by 
the effect of the inter-particle interactions in the system and vanishes in the limit
$U \rightarrow 0$.

The bulk value of the
potential in the
subsystem $B$, $V_B$ is independent of the temperature, as seen from
Fig.~\ref{1dBarrier}(a). In fact, this value is simply the thermal value
of the auxiliary field $\lambda_{th}=U\rho$. To see this, we plot
$V_B$ for systems with different values of $U$ and $\rho$ as a
function of $U\rho$ in Fig.~\ref{1dBarrier}(d), and obtain a straight
line with unit slope. 
The potential barrier across the entanglement cut increases with temperature, saturating
  at very high temperatures. This is shown in
  Fig.~\ref{1dBarrier}(e) for three different set of parameters: $U=0.5, ~ \rho =0.5$ (blue solid line),
    $U=1, ~ \rho =0.5$ (purple dashed line) and $U=1, ~ \rho =1$
    (green dashed-dotted line). The
  high temperature saturation value increases with density of the
  system as well as with the interaction in the system.

\begin{figure*}[t]
  \centering
  \includegraphics[width=0.8\textwidth]{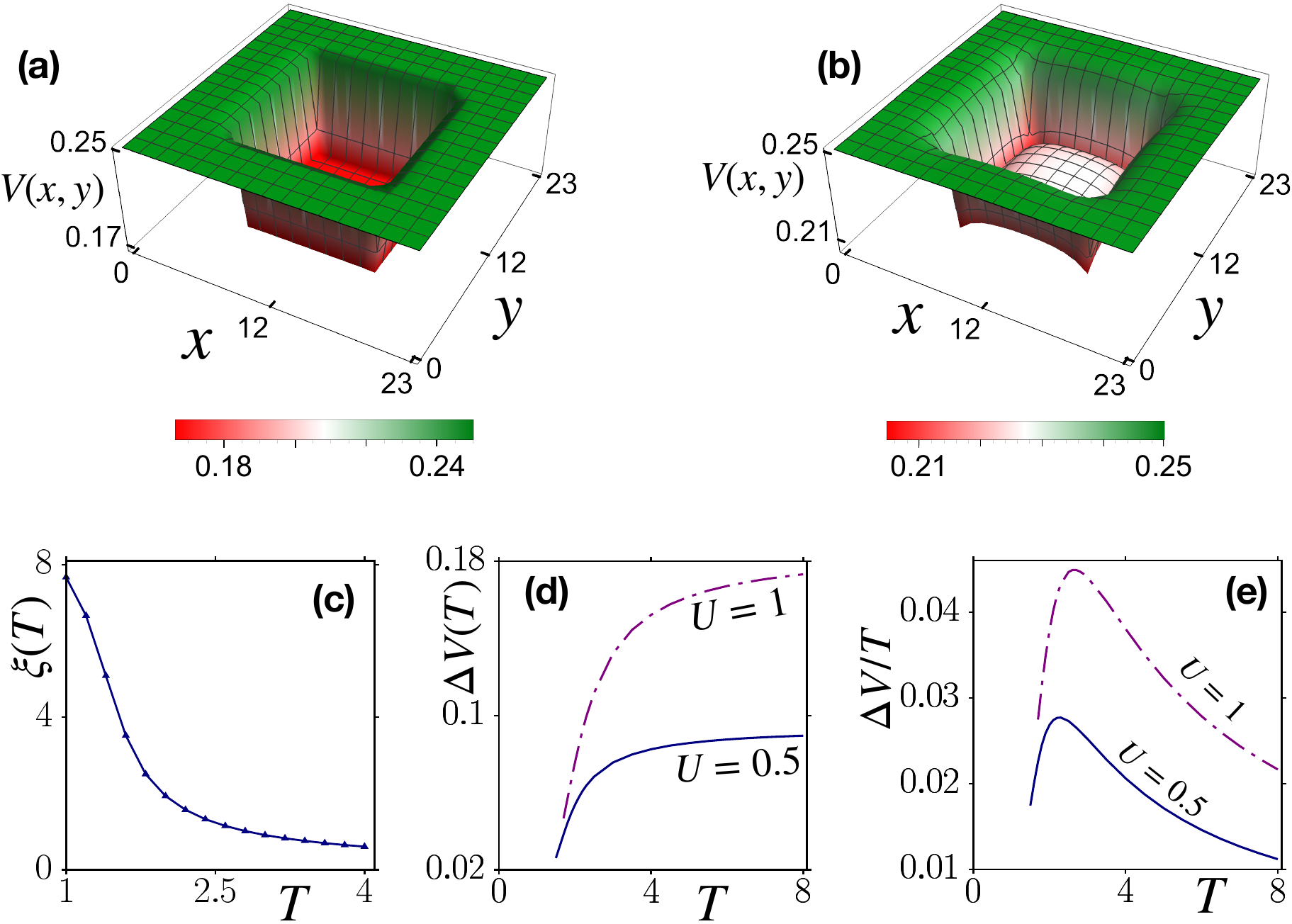}
  \caption{Profile of the effective potential $V(x,y)$ imposed by the inter-particle interaction for 2D Bosons is shown at $U=0.5$ and $\rho=0.5$ for two different temperatures: (a) for $T=4.0$ and (b) for $T=1.5$. Here, we construct $\hat{\rho}^A$ consisting of DOFs within a sub-system of size $L_A \times L_A$ located at the center of a $L \times L$ square lattice. Similar to the 1D profiles, at high $T$, $V(x,y)$ closely resembles a square well potential with $V(x,y)=V_A(T)$ inside the sub-system $A$ and a temperature independent constant $V_B=U\rho$ in the sub-system $B$. At low $T$, $V(x,y)$ deviates from the square well shape of the potential and varies across the entanglement cut on a lengthscale of the order of the thermal density-density correlation length $\xi(T)\sim 5$ at $T=1.5$. $\xi(T)$ is plotted as a function of $T$ for the same values of $U$ and $\rho$ in (c). (d)  shows the variation of $\Delta V(T)=V_B-V_A(T)$ with $T$ for $U=0.5$ (solid line) and $U=1.0$ (dashed-dotted line) for the same value of $\rho=0.5$. $\Delta V(T)$ increases with $T$ before saturating at large $T$ to a value which increases with increasing $U$. $\Delta V/T$ is plotted a as function of $T$ in (d). This shows $\Delta V/T$ reaches the peak value at an intermediate temperature scale $T=T_{\mathrm{max}}$ at which effect of inter-particle is expected to be largest in Renyi entropy. In all the plots, we use $L=24$ and $L_A=12$ and set $t=1$ and $a=1$ as units of energy and length respectively.}
  \label{veff2d}
\end{figure*}


  Since we are considering a thermal system with a potential
  barrier, the effects of this barrier, and hence of interaction in
  the system, on ${\cal S}^{(2)}$ will be controlled by the
    dimensionless parameter $\Delta V /T$. This is plotted in
    Fig.~\ref{1dBarrier}(f) as a function of temperature for three
    different set of parameters: systems with $U=0.5, ~ \rho =0.5$ (blue solid line),
    $U=1, ~ \rho =0.5$ (purple dashed line) and $U=1, ~ \rho =1$
    (green dashed-dotted line). The dimensionless parameter peaks at a
    typical value of temperature $T_{\mathrm{max}}$ before going down as $\sim T^{-1}$ at
    large temperatures. $T_{\mathrm{max}}$ increases with increasing $U$ and $\rho$.
    We should expect the effects of interaction on
    ${\cal S}^{(2)}$ to be largest around $T_{\mathrm{max}}$ .

We now focus on the Renyi entanglement entropy of the subsystems of
this interacting thermal system. The entanglement entropy scales
linearly with the size of the subsystem, as expected for a thermal
density matrix. To show this, in Fig~\ref{1dRenyi}(a), we plot the entanglement
entropy of a subsystem with $L_A$ for a system of size
$L=600$ and $U=0.5$, $\rho=0.5$ for two different temperatures,
$T=3 $ and $T=0.1 $. This allows us to define an
entanglement entropy density (entanglement entropy per site per
species of the Bosons), $\mathfrak{s}^{(2)}$ and we will now focus on
how this entanglement entropy density changes with temperature,
density and interaction in the system. In Fig.~\ref{1dRenyi}(b), we plot the entanglement entropy density for 
a subsystem of size $L_A=50$ as a function of temperature. This is done for a system with $L=100$ and
$U=1$ and two different values of density, $\rho=0.5$ and $\rho=1.0$. The entropy density increases with temperature and
saturates at high temperatures. At large $T$, $\mathfrak{s}^{(2)}$ (triangles) closely agrees with the non-interacting answer ($=\log(\coth[|\mu|/2t])$) plotted by red solid line in the inset of Fig.~\ref{1dRenyi}(b) for $\rho=0.5$. This can be understood in the following way: the chemical potential $|\mu(T)|$ (solid line with squares) increases linearly with $T$ as shown in the inset of Fig.~\ref{1dRenyi}(b) and at large $T$, $\Delta V(T)$ imposed by the inter-particle interaction becomes negligible compared to $\mu(T)$. Hence, at large $T$ the effects of interaction becomes completely suppressed in $\mathfrak{s}^{(2)}$.
This is expected
since $\Delta V/T$ goes to $0$ in this limit, as seen from
Fig.~\ref{1dBarrier}(e).

In order to understand the effect of
interaction on entanglement entropy density, we consider  $\Delta \mathfrak{s}^{(2)}=\mathfrak{s}^{(2)}(U)
-\mathfrak{s}^{(2)}(0)$ , the
difference between the entropy density of an interacting system and
that of a non-interacting system at the same temperature and
density. In  Fig~\ref{1dRenyi}(c), we plot $\Delta \mathfrak{s}^{(2)}$
as a function of temperature for a system with $\rho=0.5$ for three
different values of $U=0.5 $, $U=1.5 $ and $U=2.5 $. We note that
$\Delta \mathfrak{s}^{(2)}$ increases at low temperature and reaches a
peak before going down at large temperatures. The large temperature
decay $\sim T^{-1}$, reflecting similar decay in the dimensionless
potential barrier $\Delta V/T$, shown in Fig.~\ref{1dBarrier}(e). In
fact, the behaviour of $\Delta \mathfrak{s}^{(2)}$ is very similar to
that of $\Delta V/T$ over the full temperature range considered
here. The peak position and the peak value of $\Delta
\mathfrak{s}^{(2)}$ increases with increasing $U/t$. In
Fig~\ref{1dRenyi}(c), we have also shown the peak positions of $\Delta
V/T$ with color coded arrows for the same parameters. This shows that
the peak of $\Delta
\mathfrak{s}^{(2)}$ closely tracks that of $\Delta
V/T$.

Every density matrix of a quantum system (or subsystem) can be
interpreted in terms of an ensemble of pure quantum states drawn with
a particular classical probability distribution. If one diagonalizes the
density matrix, the eigenstates are the pure quantum states in
question and the eigenvalues are the corresponding probabilities. It
is obvious that entropy measures defined on density matrices are
actually entropy measures of the corresponding classical probabilities
associated with the density matrix. For example, the Renyi entropy of
a density matrix $S^{(2)}=-\ln \textrm{Tr} \hat{\rho}^2= -\ln \sum_i
p_i^2$, where the eigenvalues $p_i$ correspond to the classical
probability distribution.

For entanglement measures, when one
constructs the reduced density matrix of a subsystem from a pure
quantum state in the full system by tracing over degrees of freedom,
the loss of the ``quantum'' information shows up as a classical
probability distribution in the reduced description. Entanglement
entropies are various entropy measures of this distribution and
measures the classical randomness associated with the loss of the
``quantum'' information. However, when the state of the full system is
described by a density matrix, as is the case for thermal systems at
finite temperature, there is a classical randomness associated with
the full system as well. In this case, we can define the additional
randomness introduced by the loss of information (tracing), i.e. an entropy
density of tracing by considering the difference between the entropy
density of the subsystem, calculated from the reduced density matrix,
and the entropy density of the full system, calculated from the full
thermal density matrix, $\mathfrak{S}^{(2)}=\mathfrak{s}^{(2)}(L_A)
-\mathfrak{s}^{(2)}(L)$. This is a measure of the additional
randomness introduced by tracing over the subsystem $B$. Note that
since the number of degrees of freedom is different in the subsystem
and the full system, it is important to divide the entropy by the size
of the corresponding system before subtracting them.

In
Fig~\ref{1dRenyi}(d), we plot the additional entropy density
$\mathfrak{S}^{(2)}$ of a subsystem of size $L_A=50$ for a system with
$L=100$, $U=1$ and $\rho=0.5$ as a function of temperature. We see
that the additional entropy density is a monotonically decreasing
function of temperature, going to $0$ at very large temperatures. At
low temperatures, the full system is in a almost pure quantum state and 
the entropy density of the full system $\mathfrak{s}^{(2)}(L)$ is very small. Hence,
a large entropy density is gained by tracing out the degrees of freedom. On the other
hand at very high temperatures, the system behaves like a bunch of
classical free particles and hence tracing does not lead to generation
of any additional randomness in the system.

In the next section, we will extend this discussion to the case of a 2D interacting Bose gas.

\section{Renyi entropy of 2D Bose gas\label{Renyi2d}}
\begin{figure}[t]
  \centering
  \includegraphics[width=\columnwidth]{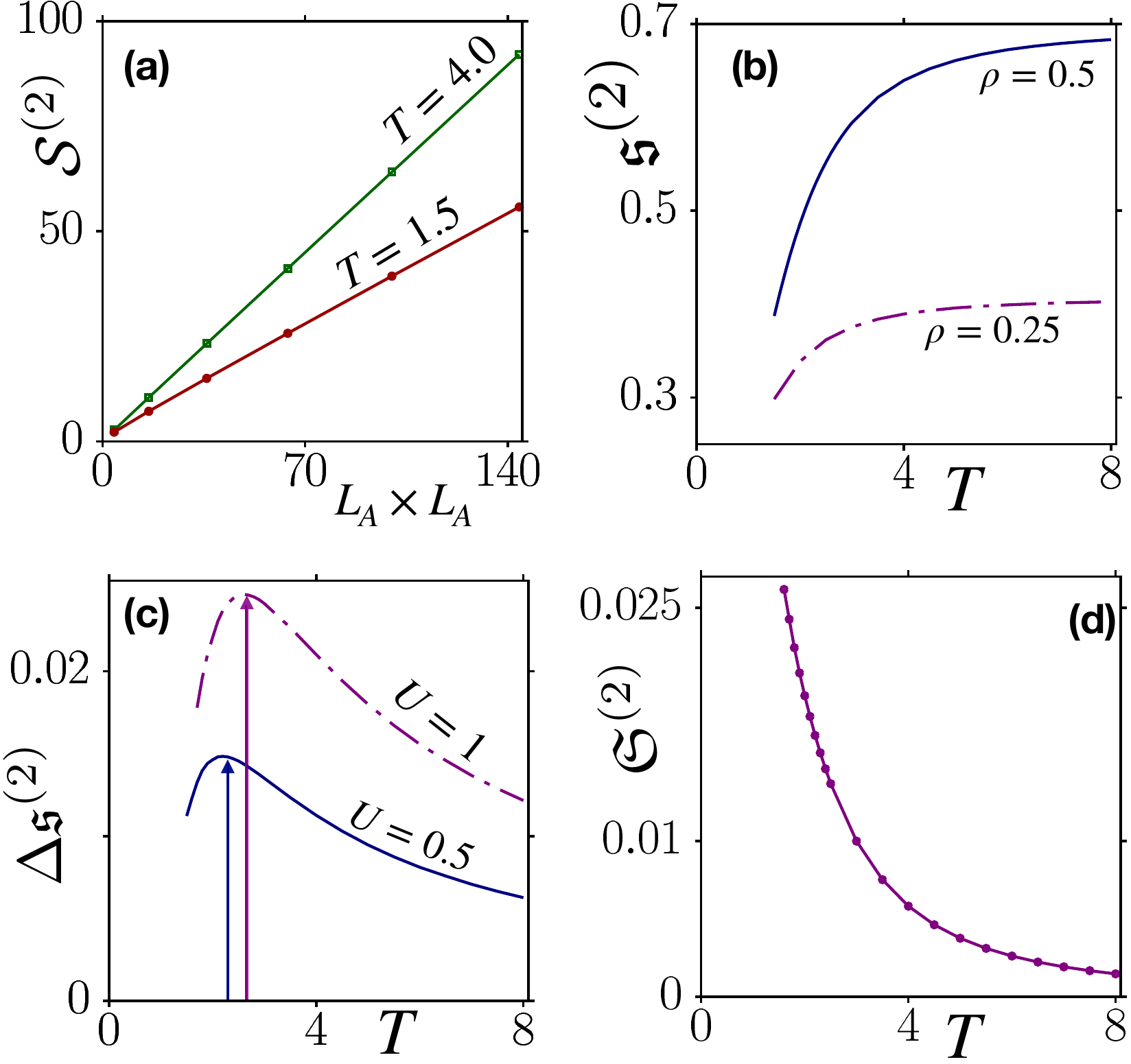}
  \caption{Renyi entropy of the sub-system $A$ for 2d Bosons: (a)
    shows linear dependence of ${\cal S}^{(2)}$ on the area, $L_A
    \times L_A$, of the sub-system $A$. The entanglement entropy
    density, $\mathfrak{s}^{(2)}$ is plotted as a function of $T$ in
    (b) for two different value of densities, $\rho=0.5$ (solid line)
    and $\rho=0.25$ (dashed-dotted line) for same value of
    $U=0.5$. $\mathfrak{s}^{(2)}$ increases with $T$ and saturates at
    high $T$ with a saturation value which increases with $\rho$. (c) shows variation of $\Delta \mathfrak{s}^{(2)}=\mathfrak{s}^{(2)}(U)
-\mathfrak{s}^{(2)}(0)$ as a function of $T$ for two different strengths of interaction: $U=1.0$ (dashed-dotted line) and $U=0.5$ (solid line) for the same value of $\rho=0.5$. This shows that $\mathfrak{s}^{(2)}$ rises with increasing $T$ and the effect of interaction becomes maximum at an intermediate temperature scale which closely agrees with $T_{\mathrm{max}}$ (shown by two solid vertical arrows) obtained from Fig. \ref{veff2d}(e). (d) shows monotonic decrease in additional entropy density $\mathfrak{S}^{(2)}=\mathfrak{s}^{(2)}(L_A)
-\mathfrak{s}^{(2)}(L)$ obtained from tracing out DOFs with increasing
$T$ for $U=0.5$ and $\rho=0.5$. 
In all the plots, we use $L=24$ and $L_A=12$.}
  \label{EE22d}
\end{figure}
In this section we extend our formalism to a $2$ dimensional
system. We would like to note that a microscopic calculation of Renyi entropy in more than one dimension is less explored in the literature and our field theory formalism which is agnostic about the dimensionality of the system is a new step in this direction. 
We consider a Bose gas on a square lattice with
nearest neighbour hopping $t$ and a local Hubbard repulsion $U$. Once
again, we consider $N$ species of Bosons and a ${\cal U}(N)$ symmetric
model given by
\bqa
H&=&-t\sum \limits_{ \langle rr'\rangle } \sum \limits_{a=1}^{N}c^{\dagger(a)}_r c^{(a)}_{r'}+h.c. \nonumber \\
 &+& \frac{U}{2N} \sum \limits_{r} \sum_{a,b=1}^{N}c^{\dagger(a)}_rc^{\dagger(b)}_r c^{(a)}_rc^{(b)}_r
\eqa
With a periodic boundary condition in both $x$ and $y$ direction, we
can work in the lattice momentum basis with a
dispersion $\epsilon_p = -2t (\cos p_x +\cos p_y)$, resulting in a bandwidth
of $8t$. The non-interacting system does not undergo a Bose Einstein
condensation at any finite temperature due to an infrared logarithmic
divergence in the number equation in $2$-d , in accordance with the
Mermin-Wagner theorem \cite{matsubara}. While the real system would show a Berezinskii
Kosterlitz Thouless type transition at a finite temperature to a disordered phase from a phase with quasi-long range
order due to proliferation of vortices, this is not captured within a
large $N$ theory, which does not account for the vortex excitations.

We consider the large $N$ approximation to the Renyi entropy of a $L_A
~\times~ L_A$ subsystem located around the center of a $L ~\times ~ L$
lattice. We will present data for
$L=24$ and $L_A=12$ in this section, unless otherwise mentioned. We first consider the
effective potential $V(x,y)$ for this system.

The effective potential
profile for a system with density $\rho=0.5$ and interaction strength
$U=0.5$ is shown in Fig.~\ref{veff2d}. Fig.~\ref{veff2d} (a) plots the
effective potential profile for a high temperature $T= 4.0$. Similar
to the one dimensional case, we find that $V(x,y)$ is almost constant in
the subsystem $B$ and sticks to its thermal value $U\rho$. The
potential forms an almost square well inside the subsystem $A$ with a
constant value $V_A$. The potentials match up at the boundary
separating the subsystem $A$ from the subsystem $B$ on a scale of the connected density-density 
correlation length. Fig.~\ref{veff2d}(b) plots the effective potential
profile for the same system at a lower temperature of $T=1.5$. In
this case, the profile varies around the entanglement cut on a larger
lengthscale. In fact, for a subsystem size of $12 \times 12$, the
potential profile within the subsystem $A$ looks almost parabolic and
leads to deep pockets at the corners of the entanglement cut. To
understand this behaviour, we have plotted the thermal connected density-density
correlation length $\xi(T)$ as a function of temperature in
Fig.~\ref{veff2d}(c). We see that the correlation length is $\sim 5$
at $T=1.5$, and hence the subsystem size is already of the same
scale as the correlation length at this temperature within our finite
size calculation. This explains the large spatial variations of the effective
potential profile in this case.

We consider the potential at the central point of subsystem $A$ to be
$V_A(T)$ and the potential at the edge of subsystem $B$ to be $V_B$ and
plot the potential barrier $\Delta V = V_B-V_A(T)$ as a function of
temperature in Fig.~\ref{veff2d}(d).  The two graphs both correspond
to a system with $\rho=0.5$. The solid line corresponds to a
system with $U=0.5$ and the dash-dotted line corresponds to a system
with $U=1$. The potential barrier increases with temperature and saturates at high temperatures to a value which increases with the interaction
strength. As mentioned before, the interaction effects are controlled
by $\Delta V/T$, which is shown in Fig.~\ref{veff2d}(e) as a function
of temperature. It shows a peak at a characteristic temperature $T_{\mathrm{max}}$, which
increases with the interaction strength. We expect the interaction
effects on entanglement entropy to be largest around $T_{\mathrm{max}}$.

We plot the Renyi entanglement entropy of the subsystem with the area of the
subsystem for two different temperatures $T=4$ and $T=1.5 $ in
Fig~\ref{EE22d}(a), which clearly shows the linear dependence of
${\cal S}^{(2)}$ with the area of the subsystem, as expected for a
thermal density matrix. The entropy per site, $\mathfrak{s}^{(2)}$ is plotted as a function
of temperature for a system size of $24\times 24 $ and a subsystem
size of $12\times 12$ in Fig~\ref{EE22d}(b). The two plots
correspond to densities of $\rho=0.5$ and $\rho =0.25$ respectively. The
interaction strength in this case is $U=0.5 $. The entropy density
increases with temperature and saturates at large temperature to
values determined by the density of the system. To understand the
effects of interaction, in
Fig~\ref{EE22d}(c) we plot $\Delta \mathfrak{s}^{(2)}=\mathfrak{s}^{(2)}(U)
-\mathfrak{s}^{(2)}(0)$, the difference between the entanglement entropy density of
an interacting system and a non-interacting system,  as a function of $T$ for a system with density
$\rho=0.5$. The interaction effects peak at an intermediate
temperature which closely follows $T_{\mathrm{max}}$ where $\Delta V/T$ is largest. This is seen from the
plots, where the location of $T_{\mathrm{max}}$ is plotted as additional
arrows. The plot also shows that the peak position increases with
increasing $U/t$. Finally, in  Fig~\ref{EE22d}(d), we plot the
difference in entropy density of the subsystem from that of  the full system $\mathfrak{S}^{(2)}=\mathfrak{s}^{(2)}(L_A)
-\mathfrak{s}^{(2)}(L)$ as a function of $T$ for $L_A=12$ and $L=24$. We see that this difference
monotonically decreases with temperature.

We note that the finite size effects are more severe in two dimensions
than in one dimension, restricting us to a regime of relatively high
temperatures. There are two reasons for this: (i) the lengthscale we
can access numerically is smaller in two dimension than in one
dimension and (ii) the infrared divergence of the number equation,
which prevents a Bose Einstein condensation, has a weak logarithmic
dependence in two dimensions compared to an inverse square root
divergence in one dimension. While this can be mitigated to some
extent by going to larger system sizes, the general trend that finite
size effects will be more severe in two dimensions will remain an
intrinsic factor for this model. 

\section{conclusion\label{Summary}}

In this paper, we have used a Wigner function based field theoretic
approach to calculate the Renyi entropy of a subsystem of interacting
Bosons in thermal equilibrium. Using a ${\cal U}(N)$ symmetric model
of lattice Bosons interacting via a local Hubbard repulsion, we derive
a functional integral for the second Renyi entropy of a subsystem of
these Bosons, when the full system is in thermal equilibrium at a fixed
density.

Using a saddle point approximation in the large $N$ limit, we
show that the entanglement entropy can be calculated in terms of an
effective system with an externally imposed potential, to be
calculated self-consistently. Although the system is translationally
invariant, the consideration of a subsystem breaks the translation
invariance. Hence the effective potential is spatially varying. We
derive the self-consistent equation for this potential and solve it
numerically for $1$ and $2$ dimensional Bose gas. In both cases we
find that the potential is flat in the bulk of subsystem $A$ and
subsystem $B$, with different values, thus creating a potential
barrier across the entanglement cut. The potential barrier increases with temperature and
saturates at high temperature. The potential varies near the boundary
between subsystem $A$ and $B$ on a scale of the density density
correlation length.

The entanglement entropy scales with the size of the subsystem (volume
law) and hence one can define an entanglement entropy density in the
subsystem. The effect of interaction on this entropy density is
largest at an intermediate temperature where the ratio of the
effective barrier to the temperature peaks. This peak temperature
increases with increase in interaction strength in the system.

There are two directions where the current formalism can be
extended. One is to go to larger system sizes, specially for higher
dimensional systems. The other is to consider fluctuations around the
static saddle point that we have considered. We leave these issues for
investigation in a future work.

\begin{acknowledgments}
The authors acknowledge the use of computational facilities at Department of Theoretical Physics, TIFR Mumbai. 
\end{acknowledgments}
\bibliographystyle{apsrev4-1} 
\bibliography{LargeN.bib}

\end{document}